# Gestão de Estoques: Modelo LEC versus Modelo (Q,R)

Cainan K. OLIVEIRA [1], Henrique G. MENCK [1], Pedro Y. TAKITO [1], Eliandro R. CIRILO [1], Paulo L. NATTI [1], Erica R. TAKANO NATTI [2]

1- Departamento de Matemática – Universidade Estadual de Londrina

2- Pontifícia Universidade Católica – Campus Londrina

**RESUMO**

Existe uma grande necessidade de se estocar materiais para a produção, porém estocar materiais tem um custo. A falta de organização no estoque pode resultar em um custo muito alto para o produto final, além de gerar outros problemas na cadeia de produção. Neste trabalho são apresentados métodos matemáticos e estatísticos aplicáveis à gestão de estoque. A análise do estoque por meio de curvas ABC dos materiais serve para identificar quais são os itens prioritários, os mais caros e com maior rotatividade (demanda). Assim é possível determinar, através de modelos de controle do estoque, o tamanho do lote de compra e a periodicidade que minimizem os custos totais de estocagem desses materiais. Por meio do modelo de Lote Econômico de Compra (LEC) e do modelo (Q,R), modelos de controle de estoque, simulou-se a minimização dos custos do estoque de uma empresa. A comparação dos resultados fornecidos pelos modelos foi realizada.

**PALAVRAS-CHAVE:** otimização; gestão de estoques; curvas ABC; Modelo LEC. Modelo (Q,R).

## 1. INTRODUÇÃO

Administração ou gestão dos materiais é uma atividade que vem sendo realizado nas empresas desde os primórdios da administração, tendo como principal objetivo atender às necessidades e expectativas dos clientes. Segundo Gonçalves (2013), no formato tradicional, a administração de materiais tem o objetivo de conciliar os interesses entre as necessidades de suprimentos e a otimização dos recursos financeiros e operacionais das empresas. Uma gestão de materiais bem estruturada permite a obtenção de vantagens competitivas por meio da redução de custos, da redução dos investimentos em estoques, das melhorias nas condições de compras mediante negociações com os fornecedores e da satisfação de clientes e consumidores em relação aos produtos oferecidos pela empresa.

Existe uma grande necessidade de se estocar materiais para a produção, porém estocar materiais tem um custo, de modo que a falta de organização pode resultar em um custo muito alto para o produto final, e a má gestão do estoque pode gerar outros problemas na cadeia de produção. Então para evitar tais custos e problemas, esse trabalho tem por objetivo trazer uma solução inteligente e otimizada para os problemas do estoque da empresa, por meios de modelos matemáticos e simulações numéricas.

Dificuldades e problemas são comuns em qualquer meio empresarial, especificamente em estoque identificamos problemas relacionados ao controle e organização dos materiais. Os problemas de organização, dimensionamento e níveis de estoque têm papel importante dentro de uma indústria, e na maioria dos casos estes problemas são tratados somente pela parte de logística. De acordo com Ching (2008), o estoque tem que ser eficiente, pois está integrado na cadeia de produção da empresa com uma função de grande importância. Nosso foco será a utilização da Matemática para descrever o desempenho do estoque e deixá-lo na sua melhor forma organizacional, funcional e rentável possível.

A empresa estudada apresenta necessidade de uma melhor gestão de seu estoque. Dois de seus principais problemas são a discordância entre o estoque físico e contábil, e a incerteza nas decisões de compras de matérias primas por oportunidade, quando estão com preços em baixa, sem levar em consideração se vale a pena comprar nesse momento, para então deixar o item estocado até o momento que ele será utilizado. Além desses, citamos outros dois problemas identificados: o acúmulo de produtos prontos parados no estoque e a falta ou excesso de material no estoque.

Na sequência apresentamos os modelos e as ferramentas matemáticas utilizadas nesse trabalho para otimizar a gestão de estoques.

## 2. METODOLOGIAS MATEMÁTICAS

Existem inúmeros modelos matemáticos e ferramentas eficientes que auxiliam as otimizações e resoluções de problemas de estoques. Nesse trabalho não estudaremos os modelos focados em organizar ou distribuir itens dentro do estoque. Tais modelos têm por objetivo facilitar o acesso pelos itens e minimizar os tempos e os custos gerados pelas movimentações dos materiais dentro do estoque, e do estoque para a produção.

Por outro lado, as ferramentas matemáticas que serão detalhadas e analisadas nos próximos capítulos serão destinadas a determinar basicamente:

- as quantidades ideais de quanto pedir de cada material,
- os momentos ideais de quando pedir de cada material, e
- os custos mínimos (ótimos) para a gestão do estoque.

Tais técnicas de controle podem ser aplicadas a todos os itens do estoque, mas assim como Ching (2008) diz, nem todos os itens estocados merecem a mesma atenção pela administração ou precisam manter a mesma disponibilidade.

### 2.1. Curva ABC

Tanto o capital empatado nos estoques como os custos operacionais podem ser diminuídos se entendermos quais são os itens mais importantes do estoque, aqueles que circulam com maior frequência e que são mais caros. Além disso, alguns itens sofrem mais concorrência em relação a outros, são mais requisitados, mais rentáveis, ou podem ter clientes que exigem um melhor nível de serviço. Por esses motivos, cada produto deve ser classificado de acordo com suas prioridades, antes de estabelecer uma política adequada de estoque. O método da curva ABC é um dos métodos que servem para esse intuito.

A curva ABC baseia-se na ideia do diagrama de Pareto, em que nem todos os itens tem a mesma importância e a atenção deve ser dada para os mais significativos. O modelo classifica os materiais em 3 faixas, e essas faixas estão relacionadas com os valores monetários de cada material no estoque, pois não basta apenas ter o preço unitário caro, mas ter pouca demanda. O valor monetário é o resultado da multiplicação entre a demanda total desse item pelo preço unitário desse material, desse modo o valor monetário de um material representa sua participação (em dinheiro) no estoque.

Por exemplo, sejam dois itens A e B. Considere que a demanda anual do item A foi de 10 unidades e o preço de uma unidade do item A é de R$ 1000,00, enquanto a demanda anual do item B foi de 100 unidades e o preço de uma unidade do item B é de R$ 100,00. Note que apesar do item A ser mais caro que o item B, ambos possuem o mesmo valor monetário, portanto são de mesma prioridade.

Para calcular a representatividade de cada item do estoque, basta calcular o valor monetário de cada item e em seguida listá-los em ordem decrescente. Finalmente deve-se calcular o percentual relativo de participação de cada item em relação ao custo total (soma de todos os valores monetários). A tabela 1 exemplifica tais procedimentos. Note que o Item1 representa 3% do valor monetário de todo o estoque, e que os itens 1, 2 e 3, juntos representam 7,75%.

| Nome Item | Demanda anual | Preço unitário | Valor Monetário | % relativo | % acumulado |
|---|---|---|---|---|---|
| Item 1 | 12000 | R$ 10,00 | R$ 120000,00 | 3,00% | 3,00% |
| Item 2 | 6667 | R$ 15,00 | R$ 100005,00 | 2,50% | 5,50% |
| Item 3 | 10000 | R$ 9,00 | R$ 90000,00 | 2,25% | 7,75% |
| . | . | . | . | . | . |
| . | . | . | . | . | . |
| . | . | . | . | . | . |
| . | . | . | . | . | 99,50% |
| . | . | . | . | 0,03% | 99,80% |
| Item 50 | . | . | . | 0,02% | 100,00% |
| total | | | R$ 4000000,00 | 100,00% | 100,00% |

Tabela 1: Cálculo da proporção monetária de cada item do estoque.

As faixas da curva ABC são classificadas em A, B e C, sendo os itens de classe A aqueles de maior prioridade (maior valor monetário). Normalmente adotam-se as seguintes proporções para cada faixa:

- Classe A: primeiros itens (da tabela 1) com maior valor monetário que acumulam até 80% desse percentual relativo de participação.
- Classe B: itens intermediários (da tabela 1) que acumulam de 80% a 95% do percentual relativo de participação.
- Classe C: últimos itens (da tabela 1) de baixo valor monetário que acumulam de 95% a 100% do percentual relativo de participação.

As proporções atribuídas para cada faixa de classe ABC podem variar de caso a caso, a situação mais comum é de 80-15-5%, citada acima. Por outro lado, em certas situações, podem ser viáveis proporções de 70-20-10%. A figura 1 apresenta uma característica recorrente da curva ABC, ou seja, que pequena parte dos itens é de classe A, enquanto que grande maioria dos itens é de classe C. Grosso modo, em geral aproximadamente 20% dos itens é responsável por 80% dos valores monetários acumulados de todos os itens.

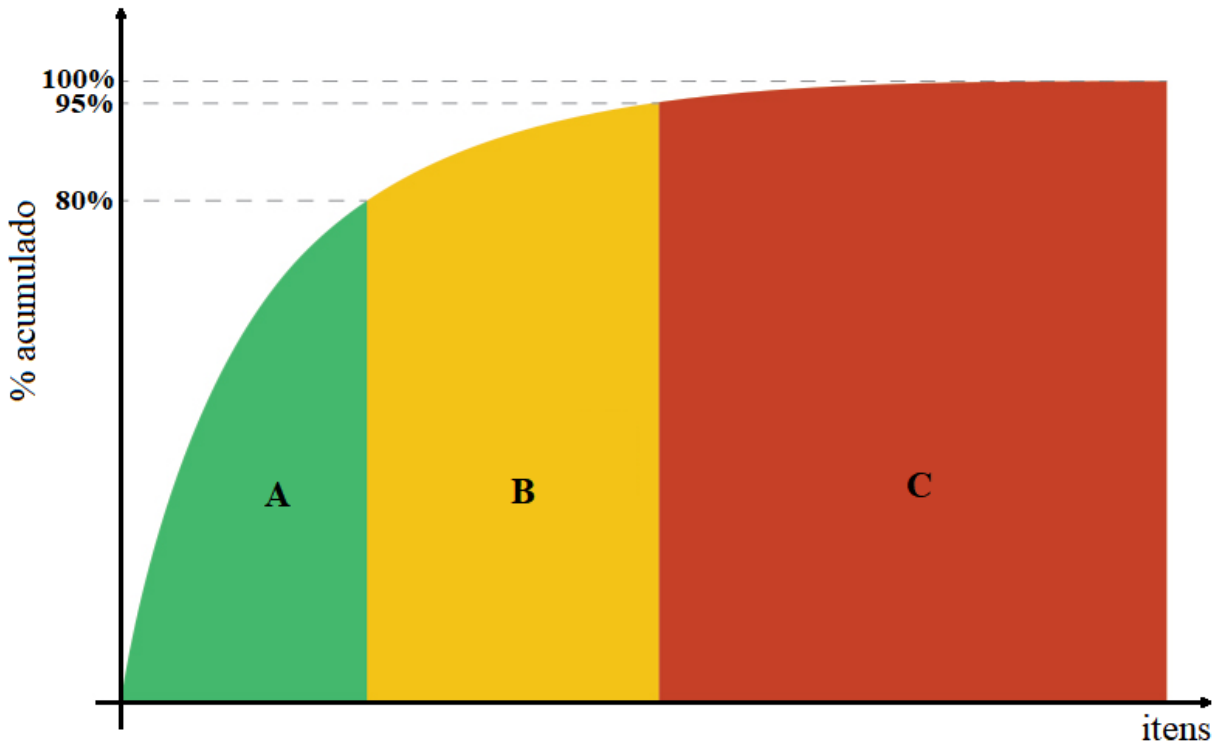

Figura 1: Curva ABC representando a quantidade de itens em cada faixa em função do valor monetário acumulado. Fonte: Próprio autor

### 2.2 Modelo LEC

Em qualquer meio empresarial tem-se o maior interesse em reduzir os custos para aumentar os lucros. Em geral, em qualquer tipo de estoque, independente de qual material esteja contido nele, os custos associados são basicamente:

a) custos de manter os estoques (custos de manutenção),
b) custos de reabastecer os estoques (custos de pedido), e
c) custos de incorrer em déficits do produto (custos de falta).

Simplificando, supondo o caso em que não ocorram déficits de produtos, o custo total gerado pelos estoques seria a soma do custo de manter e o custo de pedido. Uma questão crítica é balancear esses custos, já que eles têm comportamentos diferentes, ou seja, enquanto que o custo de manter é crescente, o custo de pedido é decrescente, quanto maior o tamanho do lote de compra. Note que quanto maiores forem as quantidades estocadas, mais espaço esses produtos ocuparam no estoque, e maiores serão os custos de mantê-lo, no entanto, quanto maior o tamanho do lote, menores serão as quantidades de reposições, então menor será a quantidade de pedidos e consequentemente menores serão os custos de pedido. As curvas na figura 2 exemplificam esses custos de estoque.

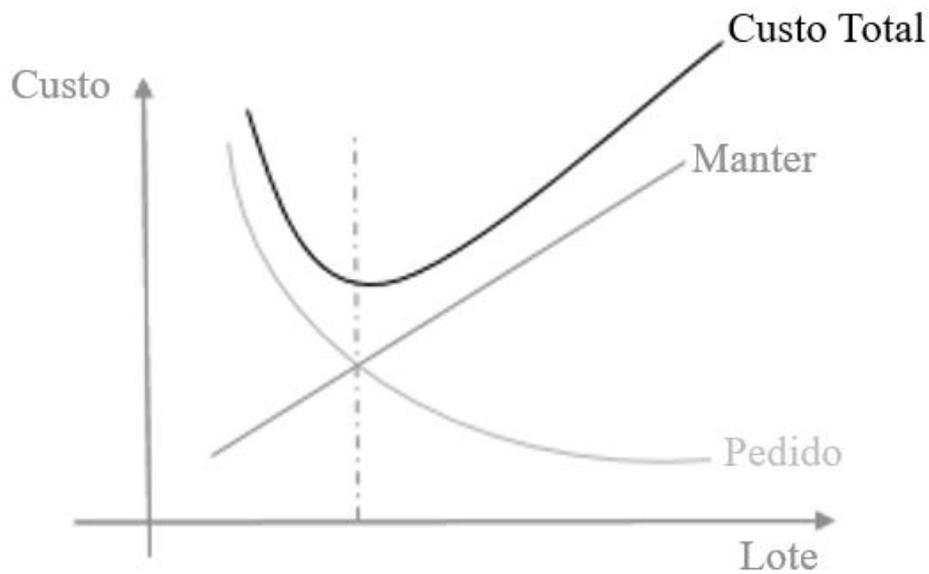

Figura 2: Comportamento dos custos gerados pelo estoque em função do tamanho do lote de compra. Fonte: Próprio autor

Observe na figura 2 que a função custo total tem o formato de uma curva com a concavidade para cima (curva convexa), o que significa que existe um valor mínimo para o custo total de estoque. Verifica-se (como veremos adiante) que esse mínimo ocorre justamente na intersecção das curvas dos custos de manter e pedir. Note que a descrição acima se torna mais complexa caso consideremos custos de déficits de produto, custo de inflação, e outros custos.

Para auxiliar o entendimento é importante definir e diferenciar o significado de demanda de um produto e o lote (de compra) do mesmo. Simplificadamente, a demanda anual de um produto pode ser suprida através de *n* lotes de compra, durante o ano. A figura 3 exemplifica essa situação.

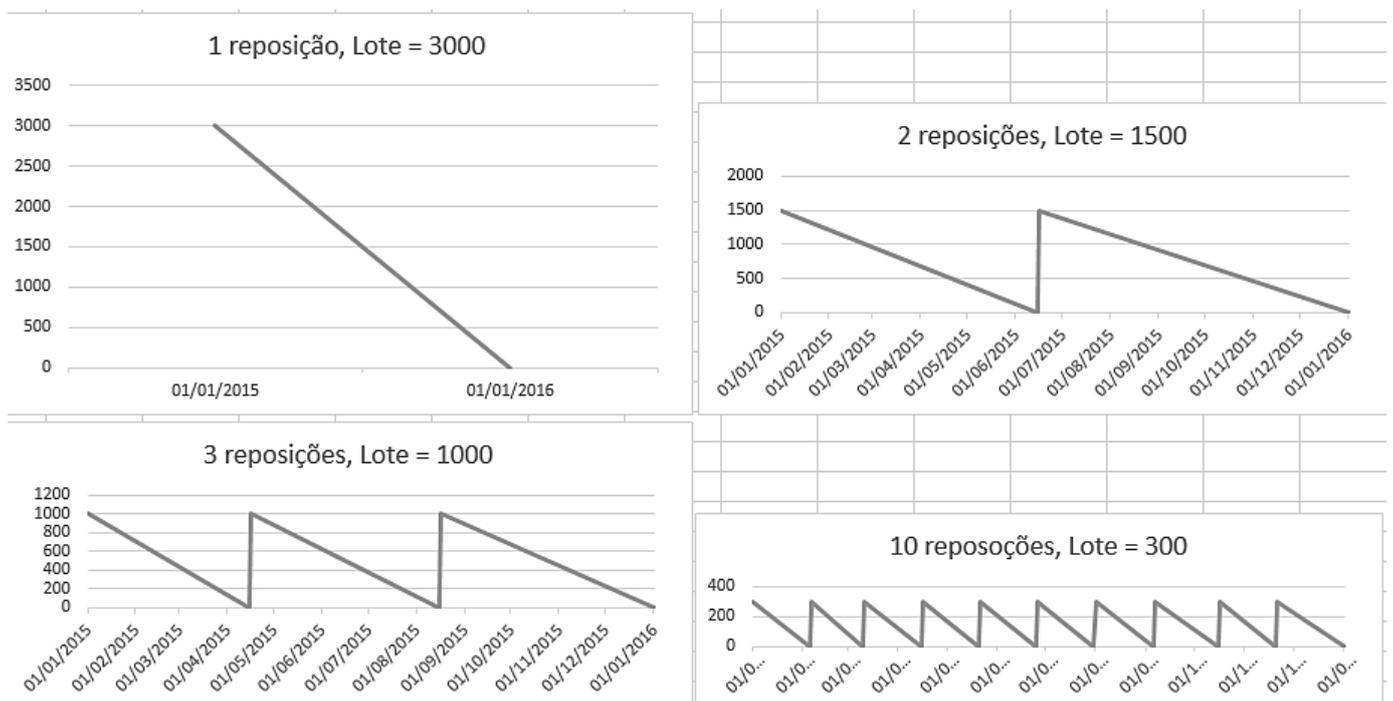

Figura 3: Produtos com mesma demanda e diferentes lotes (de compra). Fonte: Próprio autor

Na figura 3 temos o exemplo de um produto que possui uma demanda anual de 3000 unidades. Supondo que o produto tenha um nível de produção contínuo e que a taxa de utilização desse material seja constante, então note que a única coisa que varia são as quantidades de reposições e o tamanho do lote. Basicamente, demanda significa a quantidade que foi requisitada pela produção desse determinado item, e o lote é a quantidade máxima que será estocada desse item em cada reposição, lembrando que em situações mais comuns o tamanho do lote é diferente em cada reposição, mas de qualquer maneira a soma de todos os lotes tem que ser, no mínimo, igual à demanda.

Então, temos como um dos objetivos determinar quais são as quantidades ideais de reposição e o tamanho do lote, de cada material, que minimizem os custos gerados dentro do estoque, mas sem deixar de lado o objetivo principal da gestão de estoque que é de dar garantia do suprimento dos materiais necessários ao bom funcionamento da empresa, evitando faltas, paralisações eventuais na produção e satisfazendo às necessidades dos clientes e usuários.

O Modelo do Lote Econômico de Compra (LEC) é um modelo básico de controle de estoque, que permite determinar uma quantidade ótima de pedido de compra para um item do estoque, tendo em vista minimizar os custos totais de estocagem (por isso a denominação de Lote Econômico). Definindo as seguintes quantidades abaixo:

*Q*: tamanho do lote de compra,

*D*: demanda anual do produto,

*P*: Preço de compra unitária,

*Cm*: Custo unitário de armazenagem,

*Cp*: Custo unitário do pedido,

a modelagem matemática dos custos associados ao estoque, de acordo com modelo LEC, é dada abaixo:

- Custo de manutenção: $\frac{Q}{2} * C_m$ \hfill (1)

  São os custos diretamente proporcionais à quantidade estocada. Incluem os custos de armazenagem, os custos de seguro, os custos de manuseio, os custos de obsolescência, etc.

- Custo de pedido: $\frac{D}{Q} * C_p$ \hfill (2)

  São os custos inversamente proporcionais à quantidade estocada. Note que a demanda dividida pelo tamanho do lote representa a quantidade do item reposicionado, ou seja, quantidade de pedido. O custo associado ao trabalho de efetuar o pedido de determinado lote de produtos engloba custos de mão de obra, de transporte de pedido, controle do recebimento do produto, controle de qualidade do pedido recebido, entre outros. No caso de itens fabricados são chamados de custos de preparação.

- Custo de aquisição: $D * P$ (3)

Correspondem aos custos de compra dos materiais que serão estocados, ou seja, é o valor monetário do lote, demanda multiplicada pelo preço unitário.

Portanto, a função objetivo do modelo LEC pode ser expressa como:

$$\min = f(Q) = \frac{Q}{2} * Cm + \frac{D}{Q} * Cp + D * P \quad (4)$$

Como vimos na figura 2 acima, a função do custo total $f(Q)$, que queremos minimizar, tem concavidade para cima. Para encontrar o valor de Q que minimize o custo total, basta derivar $f(Q)$ em relação a Q, igualar a zero. Isolando Q, obtendo assim o valor do lote econômico ótimo, ou seja:

$$Q = \sqrt{\frac{2*D*Cp}{Cm}} \quad (5)$$

Note que podemos obter o resultado (5) igualando os custos de manutenção (1) com o de pedido (2), de modo que o custo mínimo ocorre como apresentado na figura 2, isto é, na intersecção das curvas de custo de manutenção e de pedido.

### 3. RESULTADOS

Considere a tabela 2 que apresenta a movimentação de produtos de uma empresa. Ela fornece o tipo/nome do item; o tipo do movimento ("E" se entrou, "S" se saiu); a data do movimento; a quantidade (em unidade) e a quantidade (em KG), positiva se entrou, negativa se saiu.

| Nome | Movimento | Data | Qtd. | Saldo KG |
|---|---|---|---|---|
| item A | E | 10/01/2011 | 200 | 22,8 |
| item A | S | 24/09/2011 | -300 | -34,2 |
| item B | E | 10/01/2011 | 734 | 88,08 |
| item B | E | 31/03/2011 | 3352 | 402,24 |
| Item C | E | 17/08/2011 | 1500 | 265,5 |
| item C | E | 20/10/2011 | 300 | 531 |
| item C | S | 30/08/2011 | -692 | -122,484 |
| item C | S | 31/08/2011 | -682 | -120,714 |
| item D | E | 14/01/2011 | 400 | 12 |
| item D | S | 05/01/2011 | -60 | -1,8 |

Tabela 2: Exemplo de parte da movimentação do estoque em uma empresa.

Se os preços são referentes a cada quilograma do material, então conhecendo os preços por quilo de cada material, obtemos a classificação de valor monetário dos materiais pela curva ABC. Identificados os itens de classificação A, que representam em valores monetários 80% do estoque, passemos a estudá-los. Os itens classificados como B e C não serão considerados na análise do estoque. Portanto, dos itens considerados como de classe A e dos dados da tabela 2, pode-se construir gráficos que representam a movimentação desses itens nos últimos anos para observar se eles possuem frequências de rotatividade estável, se é periódica, como é o tipo de movimentação do material, suas quantidades máximas de lote de compra e se é gerenciado de forma eficiente. Seguem alguns exemplos de gráficos de movimentação de produtos em estoques.

### 3.1. Tipos de movimentação de produto no estoque

Na figura 4 vemos um exemplo de gráfico de movimentação de um produto do estoque, denominado item A, que apresenta certa homogeneidade nas flutuações, sem muitos picos de grande amplitude e com movimentações periódicas.

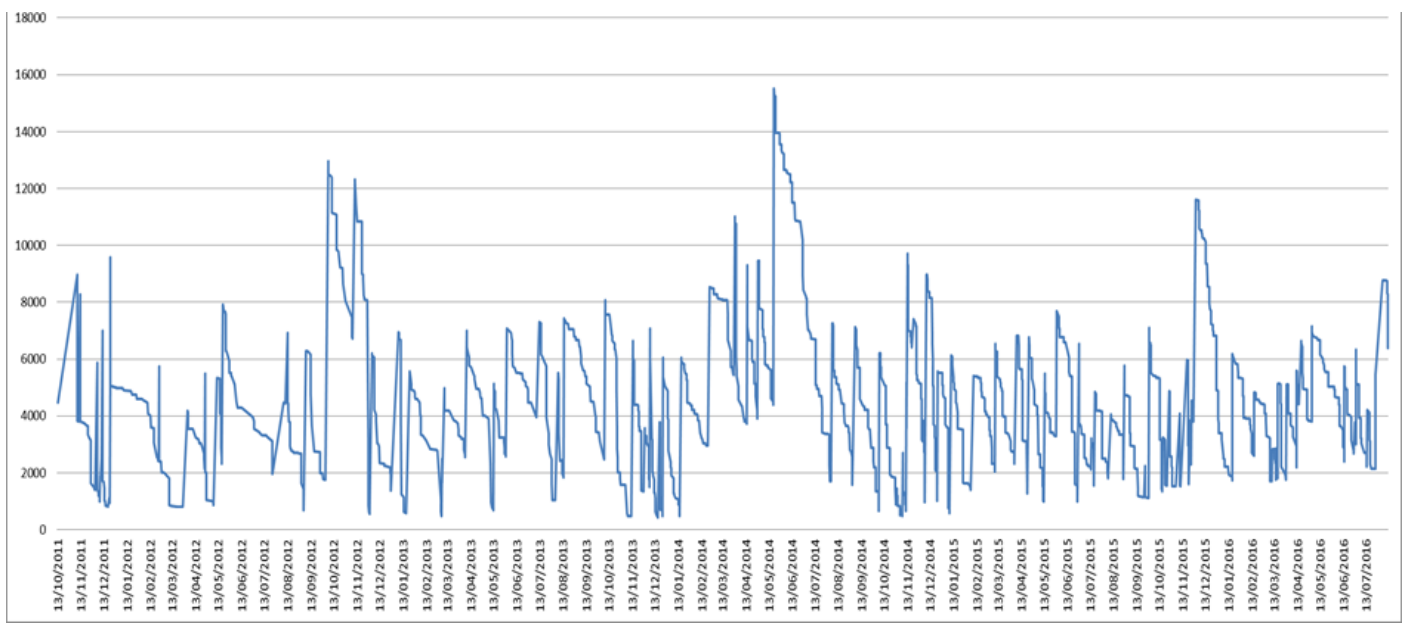

Figura 4: Gráfico da movimentação do item A no estoque. Fonte: Próprio autor

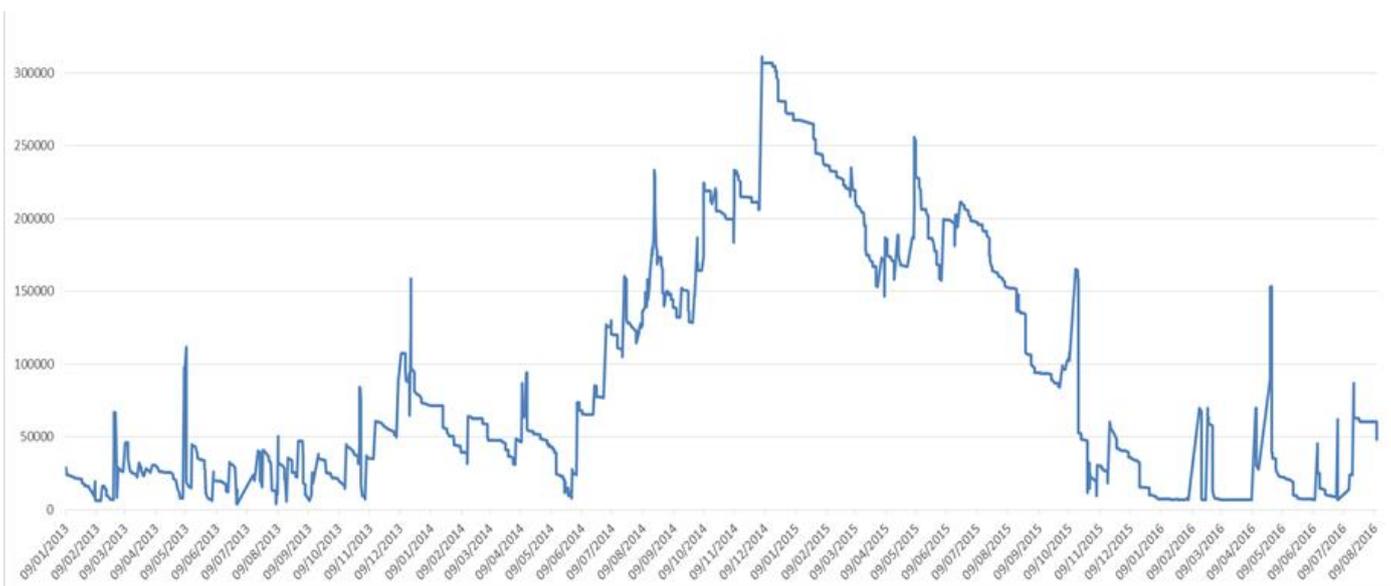

Figura 5: Gráfico da movimentação do item B no estoque. Fonte: Próprio autor

Por outro lado, na figura 5, observamos o gráfico de movimentação de um produto do estoque, denominado item B, que apresenta um acúmulo de entradas (compras) no período entre maio de 2014 a outubro de 2015. Nesse período certa quantidade desse material esteve parada no estoque, e mesmo assim houve compras, provavelmente compras de oportunidade, como aquela que ocorreu em dezembro de 2014. Tais situações são designadas como patológicas, gerando custos elevados de estocagem. Em tais circunstâncias o modelo LEC pode propor uma gestão de estoque mais econômica.

Também existem casos de itens com gráficos de movimentação como apresentado na figura 6, onde observamos que um produto, denominado item C, é comprado e já utilizado, logo em seguida. Essa movimentação, do tipo *Just In Time*, faz com que os itens fiquem pouco tempo parado no estoque, o que é bom. A movimentação *Just In Time* é um modelo que visa utilização mínima do estoque e já é ótima, consequentemente o modelo LEC não deve ser aplicado para tais itens.

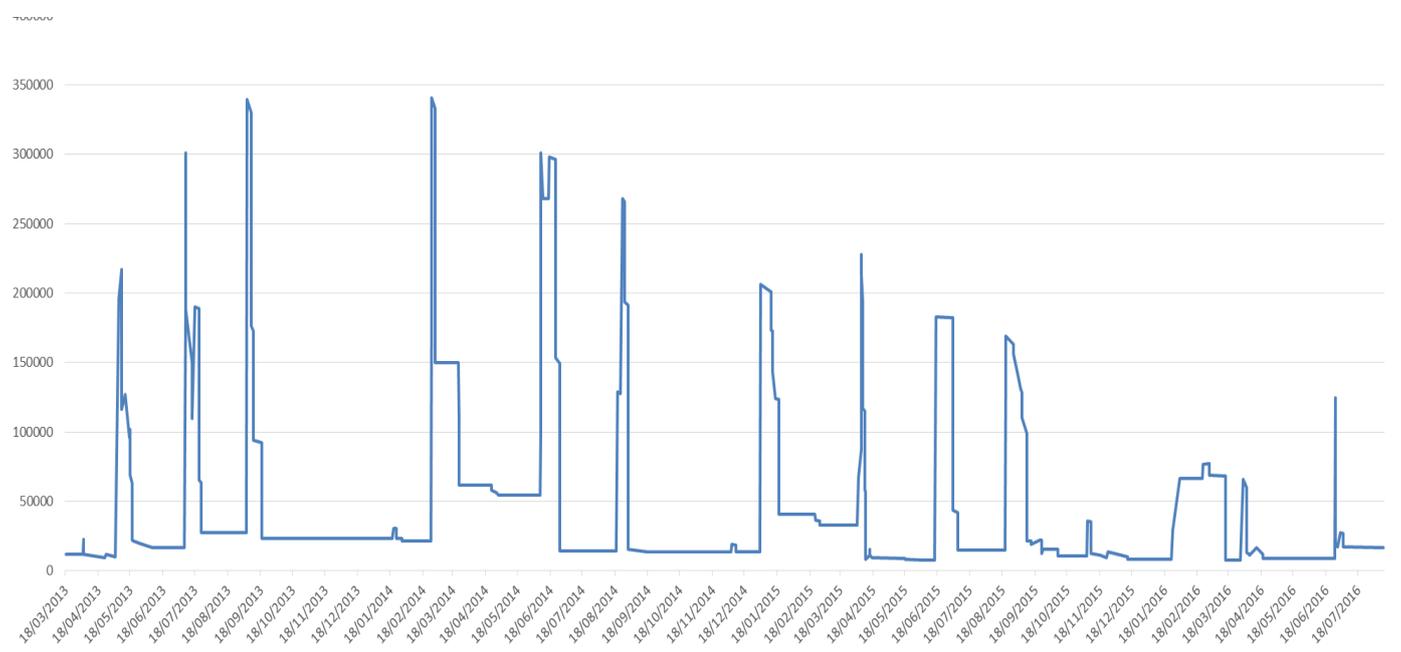

Figura 6: Gráfico da movimentação do item C no estoque. Fonte: Próprio autor
.

### 3.2. Cálculo das quantidades *Cm* e *Cp* para os itens do estoque

Para aplicar o modelo LEC precisamos conhecer quais são os valores dos parâmetros *Cm*, custo unitário de armazenagem, e *Cp*, custo unitário do pedido, em *f(Q)* dada em (4).

Sejam os custos anuais de armazenagem, denotados por *CM*, os custos com pessoal (folha de pagamento e encargos trabalhistas), custos de manutenção do estoque (aluguéis, impostos, seguros, manutenções,...), custos de escritório (papéis, materiais, impressoras,...), custos de movimentação (empilhadeiras, combustíveis,...), custos gerais (água, luz, telefone,...), entre outros.

Sejam os custos anuais de pedido, denotados por *CP*, os custos com pessoal (folha de pagamento e encargos trabalhistas), custos de escritório (papéis, materiais, impressoras,...), custos de transporte (combustível, frete, pedágio,...), custos gerais (água, luz, telefone,...), entre outros.

Quantificados os gastos totais anuais de armazenagem *CM* e os gastos totais de pedido *CP*, podemos calcular as quantidades unitárias *Cm* e *Cp*. Modelando o cálculo matemático dessas quantidades propomos que elas sejam dadas pelas equações abaixo:

$$Cm = CM/A_t \quad , \quad Cp = CP/E_t \tag{6}$$

onde $E_t$ é a quantidade de reposição de todos os itens somados, enquanto $A_t$ é a soma de todas as áreas dos gráficos de movimentação de todos itens considerados.

Tendo calculado os valores para as quantidades unitárias *Cm* e *Cp*, basta aplicá-las no modelo LEC e teremos o valor para o lote econômico de qualquer item que desejarmos do estoque.

### 3.3. Verificação da constância da taxa de consumo dos itens do estoque

Identificado os itens prioritários de classe A e calculado as quantidades *Cm* e *Cp* do modelo LEC, deve-se verificar, por exemplo, através de uma análise gráfica, se a taxa de consumo dos itens selecionados é constante, ou quase constante, uma das premissas do modelo LEC.

Construindo gráficos de movimentações de entradas e saídas dos itens de classificação A, e gráficos de taxas de consumo dos itens de classe A, é possível notar alguns diferentes comportamentos entre os itens. Seguem alguns exemplos.

Na figura 7 apresentamos a movimentação durante 2016 de um item de classe A, denominado item D. Observa-se uma movimentação com certa homogeneidade nas flutuações, sem muitos picos de grande amplitude e com movimentações periódicas. Na figura 8 apresentamos a taxa de consumo anual do item D, comparando-a com uma reta (consumo constante). Para construir esse gráfico, somamos todas as entradas no ano e começamos o gráfico com essa condição inicial, em 04/01/2016. Depois, sucessivamente, subtraímos todas as saídas esse item D durante 2016.

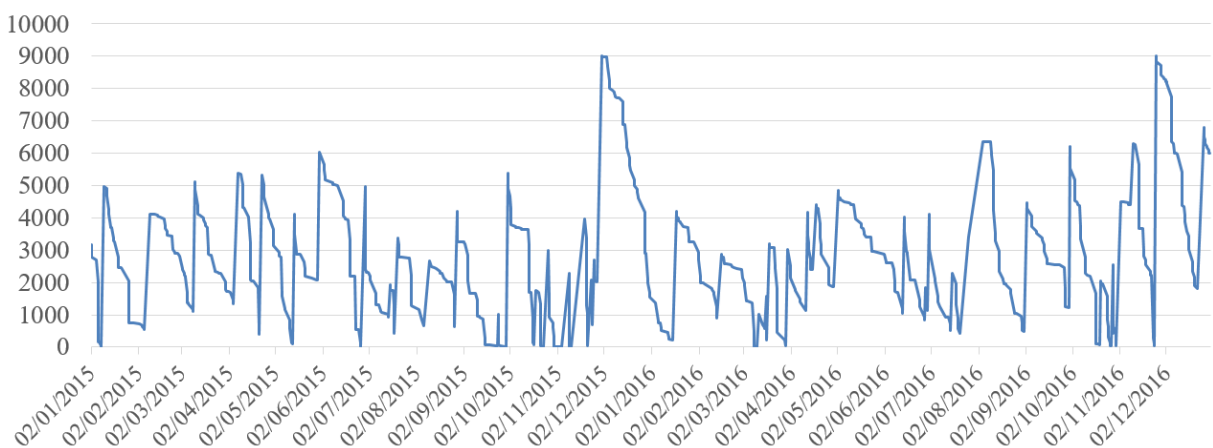

Figura 7: Gráfico de movimentação do item D durante 2016. Fonte: Próprio autor

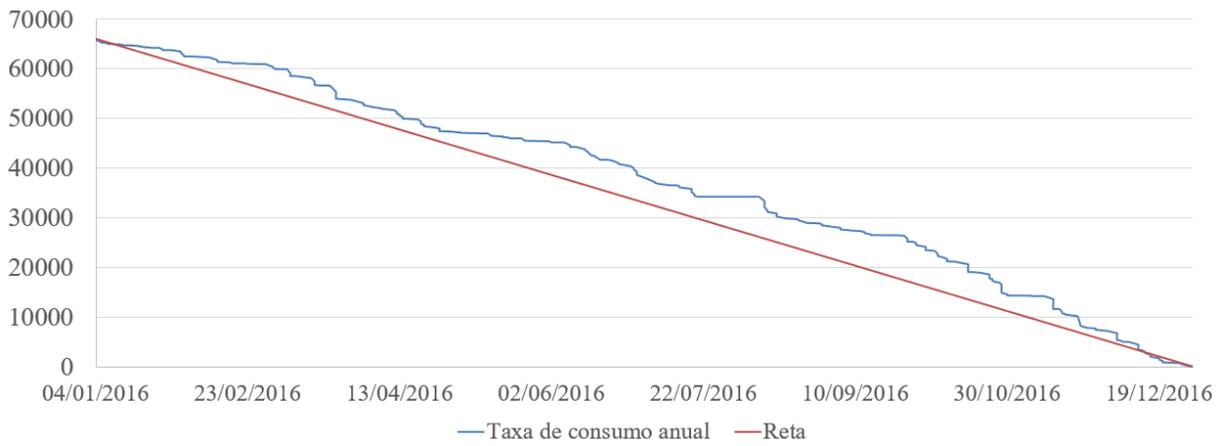

Figura 8: Gráfico da taxa de consumo do item D durante 2016 (curva azul) comparado com uma taxa de consumo constante (reta vermelha). Fonte: Próprio autor

Observe na figura 8 que a taxa de consumo não é constante (hipótese de vários modelos de gestão de estoque), mas que está próxima da reta traçada. Note que a taxa de consumo (demanda) anual do item D em 2016 foi de 65000 kg, o que permite estimar um consumo médio de 177,60 quilos, por dia, desse item.

Sob determinadas circunstâncias é comum uma empresa realizar uma compra de oportunidade, inchando o estoque com um produto. Tais situações evidenciam a possibilidade de melhoria na gestão do estoque. Na figura 9 apresentamos a movimentação durante 2016 de um item de classe A, denominado item E.

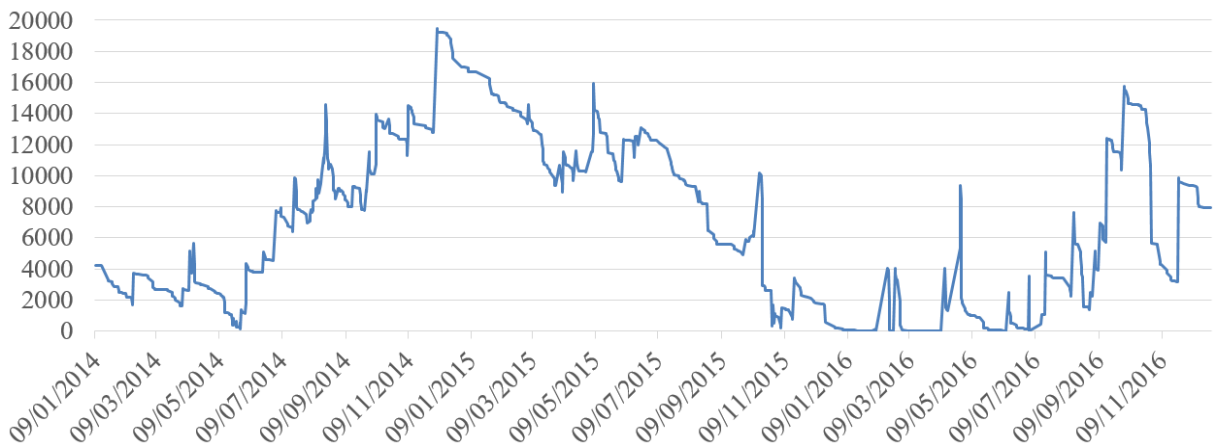

Figura 9: Gráfico de movimentação do item E durante 2016. Fonte: Próprio autor

Na figura 9 observe-se que no período entre junho de 2014 a outubro de 2015 existiu excesso do item E no estoque. Nesse período, por alguma razão, o gestor continuou comprado lotes do item E, mesmo com a existência de grandes quantidades desse item já estocadas. Obviamente, esse estoque inchado gerou custos extras. Por outro lado, quando observamos a figura 10, notamos que a taxa de consumo do item E nesse período se mostra praticamente constante, o que deixa evidente a possibilidade de melhoria na gestão do estoque para esse item.

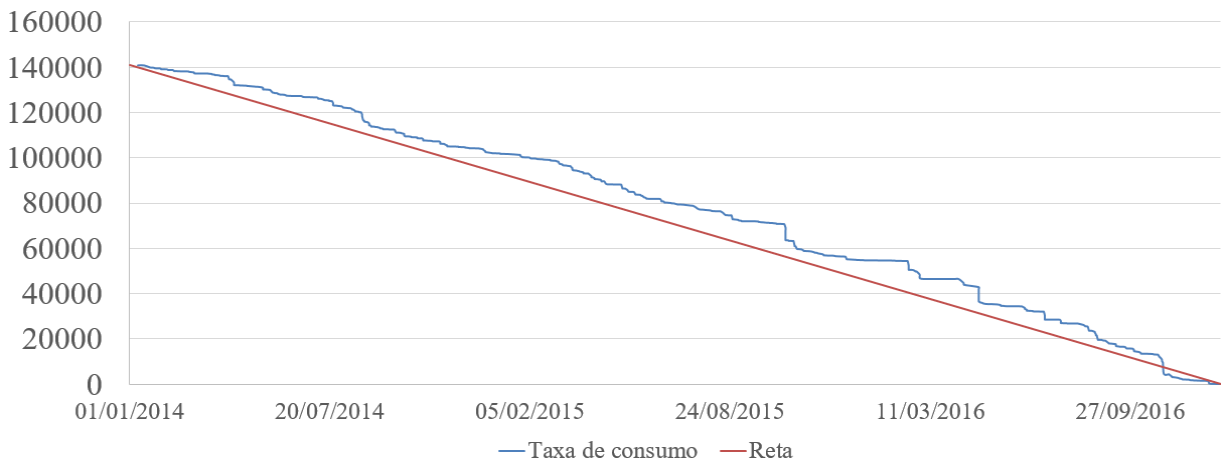

Figura 10: Gráfico da taxa de consumo do item D durante 2016 (curva azul) comparado com uma taxa de consumo constante (reta vermelha). Fonte: Próprio autor

### 3.4. Aplicações aos modelos LEC e (Q, R)

A aplicação pode ser realizada a qualquer item do estoque que satisfaça as hipóteses dos modelos. Aplicaremos o modelo LEC e o modelo (Q,R) a um mesmo item de classificação A para comparar os resultados obtidos.

O item selecionado será denominado de item $\beta$ e o gráfico de movimentação de entradas e saídas do item $\beta$ possui o comportamento apresentado na figura 11. Note que a movimentação desse item durante 2016 apresenta três grandes lotes econômicos de compra de aproximadamente 9000 unidades, e vários outros lotes econômicos de compra menores de até 1000 unidades. A primeira questão que surge é se esses tamanhos de lotes são realmente economicamente eficientes. A partir da análise realizada pelos modelos de gerenciamento de estoques será possível obter essa resposta.

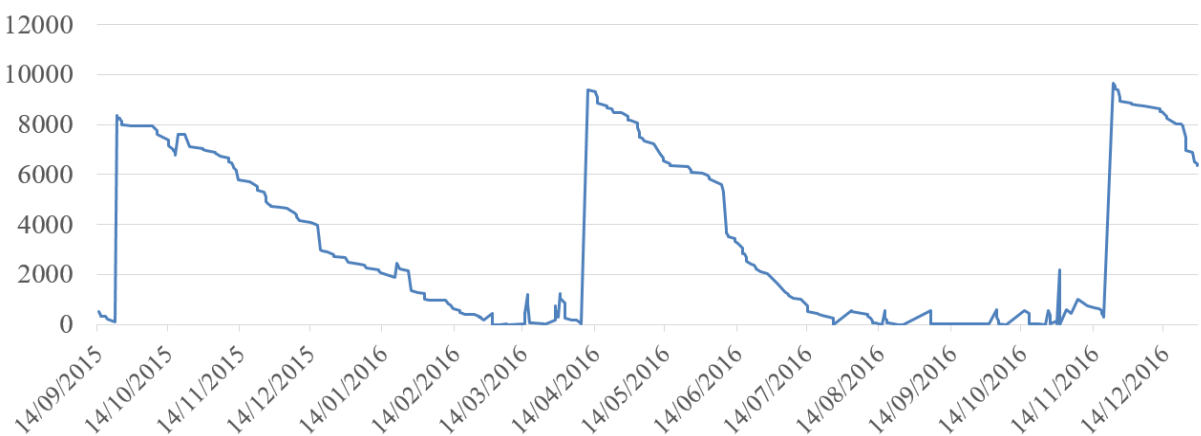

Figura 11: Gráfico de movimentação do item β durante 2016. Fonte: Próprio autor

A seguir analisemos algumas informações importantes referentes ao item β. Primeiramente, o item $\beta$ precisa apresentar uma taxa de consumo constante ao longo do tempo, hipótese do modelo LEC. Veja na figura 12 a taxa de consumo anual do item $\beta$ ao longo de 2016.

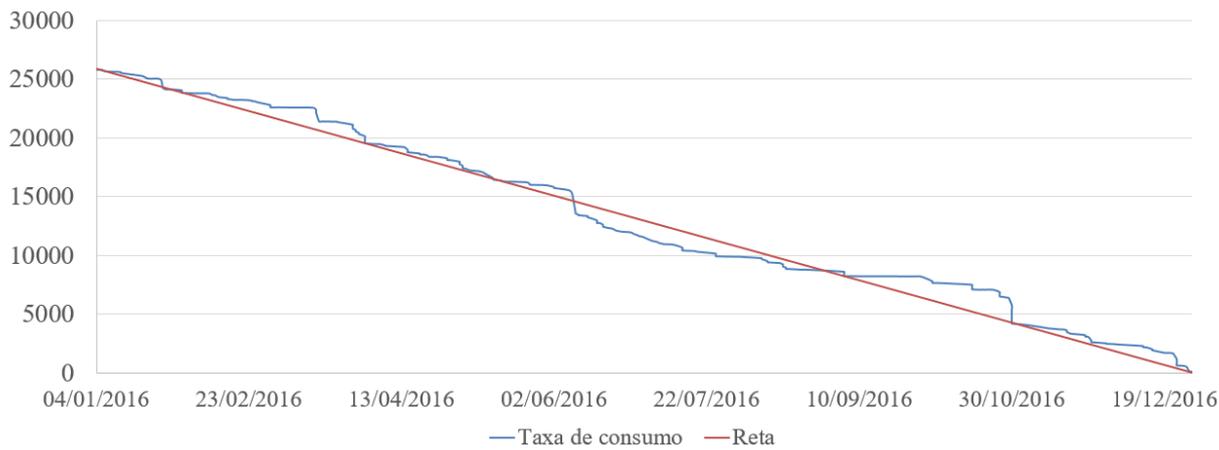

Figura 12: Gráfico da taxa de consumo do item β durante 2016 (curva azul) comparado com uma taxa de consumo constante (reta vermelha). Fonte: Próprio autor

Note que a taxa de consumo do item *β* se manteve praticamente constante durante o ano de 2016. Para uma demanda anual de 25550 unidades, a taxa de consumo diário do item *β* é de 70 unidades por dia.

Sobre o *lead-time,* ou prazo de entrega, varia de item para item e de fornecedor para fornecedor, pois podem existir itens que o fornecedor demore mais para processar, ou pode ser que sejam itens importados de outros países. No caso do item *β* o *lead-time* é conhecido e é de 14 dias.

### 3.4.1 Resultados do modelo LEC

Todas as informações necessárias para poder aplicar o modelo LEC foram obtidas. Considere os parâmetros apresentados na tabela 3.

| D | Cm | Cp |
|---|---|---|
| 25550 | R$ 1,00 | R$ 50,00 |

Tabela 3: Parâmetros do item β para o modelo LEC. Fonte: Próprio autor

Substituindo as informações nas equações do modelo LEC, obtém-se que:

$$Q_{LEC} = \sqrt{2 * 25550 * 50,00/1,00} \sim 1600 \text{ unidades.} \tag{7}$$

Portanto, para atender a demanda no período de 2016, o tamanho do lote ideal de compra, que teria minimizado os custos gerados pelo estoque, seria de aproximadamente 1600 unidades.

Quanto ao momento de pedido, como a taxa de consumo diário é de 70 unidades por dia e o *lead-time* é de 14 dias, esse pedido deve ser realizado quando o lote atingir a seguinte quantidade:

Ponto de encomenda = 70 ∗ 14 = 980 unidades.                               (8)

Deseja-se agora calcular qual foi a margem de lucro obtida com esse resultado. Para isso é preciso saber quais foram os gastos com armazenagem e os gastos com pedidos do item *β*. Tais informações estão na tabela 4 e foram extraídas da movimentação do item *β* apresentada na figura 11, ou seja,

|  | Estoque médio | Total de pedidos |
|---|---|---|
| Dados | 2580 unidades | 24 |

Tabela 4: Dados obtidos da movimentação do item β durante 2016. Fonte: Próprio autor

Aplicando na fórmula do custo de manutenção e pedido de estocagem, equação (4), segue que o custo anual real de estocagem do item *β* durante 2016 foi de :

Custo de estocagem = 2580 ∗ 1 + 24 ∗ 50 = R$ 3780,00                       (9)

Logo, apenas com armazenagem e pedidos foram gastos R$ 3780,00 com esse item durante 2016.

Agora vejamos quanto teria sido gasto caso tivéssemos comprado lotes de 1600 unidades, como proposto pelo modelo LEC:

$$\text{Custo Total LEC} = \frac{1600}{2} * 1 + \frac{25550}{1600} * 50 \sim \text{R\$ } 1600,00 \qquad (10)$$

Note que aplicando o modelo LEC somente no item *β*, a empresa poderia ter lucrado R$ 2180,00. Salientamos que os custos de manutenção e de pedido não diminuem linearmente com a diminuição do estoque. Obviamente o lucro, ou a economia, com a utilização do modelo LEC seriam menores do que esse mensurados pelo modelo LEC.

### 3.4.2 Resultados do modelo (Q,R)

Sobre as informações necessárias para a implementação dos cálculos para o modelo (Q,R) considere os dados apresentados na tabela 5.

| D | Cm | Cp | Cf | *lead-time* |
|---|---|---|---|---|
| 25550 | 1,00 | 50,00 | 4,00 | 14 dias |

Tabela 5: Parâmetros do item β para o modelo (Q,R). Fonte: Próprio autor

onde Cf é o custo de escassez e ocorre quando há o esgotamento de um item do estoque. Substituindo os valores na equação do valor do lote ideal de encomenda do modelo (Q,R), segue que o lote ideal para o item β é dado por (Hillier; Lieberman, 2013)

$$Q^* = \sqrt{\frac{2*25550*50,00}{1,00}} \sqrt{\frac{4,00+1,00}{4,00}} \sim 1800 \text{ unidades.} \tag{11}$$

Para determinar o ponto de encomenda ideal é necessário extrair mais algumas informações sobre o item *β*, como por exemplo, a média $\mu^*$ e a variabilidade $\sigma^{2^*}$ do consumo do período de 2016 e escolher a probabilidade desejada pela gerência para que não ocorra esgotamento de estoque. Sejam os dados apresentados na tabela 6.

| L | $\mu^*$ | $\sigma^{2^*}$ |
|---|---|---|
| 0,75 | 981,75 | 520143,86 |

Tabela 6: Outros dados necessários para o modelo (Q,R). Fonte: Próprio autor.

Com essas informações e admitindo que a probabilidade de não ocorrer esgotamento seja de 75%, obtém-se que

Ponto de encomenda ~1400 unidades. (12)

Comparando os resultados obtidos do modelo (Q,R), levando em conta a probabilidade desejada de não ocorrer esgotamento de 75%, com os resultados obtidos anteriormente pelo modelo LEC, seguem os resultados na tabela 7.

|  | Lote ideal de compra | Ponto de reposição |
|---|---|---|
| LEC | 1600 | 980 |
| (Q,R) | 1800 | 1400 |

Tabela 7: Comparações dos resultados dos modelos LEC e (Q,R). Fonte: Próprio autor

Observe que o resultado obtido para o tamanho do lote de compra pelo modelo (Q,R) é maior do que o obtido para o modelo LEC. Analogamente, o ponto de reposição obtido pelo modelo (Q,R) também é maior do que o ponto de reposição do LEC. O motivo disso pode se resumir em uma palavra, "precaução". Diferente do modelo LEC, o modelo (Q,R) leva em conta os custos de escassez e a probabilidade de ocorrerem faltas no estoque, o que é mais conveniente e fornece maior garantia em casos de emergências e imprevisibilidades.

## 4. CONCLUSÃO

Nesse trabalho foi mostrado os procedimentos para otimização de um estoque de uma empresa. Num estudo de caso de um item específico de um estoque se verificou margens de economia de até 50% na redução nas despesas do estoque. Salientamos que os custos de manutenção e de pedido não diminuem linearmente com a diminuição do estoque, como previsto no modelo LEC.

Normalmente, na gestão de estoques, observa-se que as movimentações no estoque são aleatórias e as compras são superdimensionadas, devido principalmente às compras por oportunidade.

Nesse contexto, caso as demandas futuras sejam previsíveis, o modelo LEC fornecerá resultados que irão minimizar os custos totais de estocagem. Por outro lado, em nosso contexto de uma economia instável, é mais viável a utilização do modelo (Q,R), que fornecerá uma maior garantia de gerenciamento de estoques.

## 5. AGRADECIMENTOS

## 6. REFERÊNCIAS

BOWERSOX, D. J.; CLOSS David J.; COOPER M. Bixby; BOWERSOX, John C.; Gestão Logística da cadeia de suprimentos. Porto Alegre: AMGH Ed., 2014.

CHING, H. Y.; Gestão de Estoque na Cadeia de Logística Integrada: Supply chain. São Paulo: Ed. Atlas, 2008.

COELHO, Leandro C. Entendendo o Lote Econômico de Compras. Disponível em: <http://www.logisticadescomplicada.com/entendendo-o-lote-economico-de-compras-lec-ou-eoq/> Acesso em 20 out 2016.

ELLENRIEDER, A. V.; Pesquisa Operacional. Rio de Janeir : A. Neves, 1971.

GONÇALVES, P. S.; Administração de Materiais. Rio de Janeiro: Elsevier, 2013.

HILLIER, F. S.; LIEBERMAN, G. J.; Introdução à pesquisa operacional, 9.ed. Porto Alegre: AMGH Ed., 2013.

MORETTIN, P. A.; BUSSAB, W. O. Estatística básica, 8.ed. São Paulo: Ed Saraiva, 2013.

MOURA, C. E.; Gestão de Estoque: Ação e Monitoramento na Cadeia de Logística Integrada. Rio de Janeiro: Ciências Moderna, 2004.

POZO, H.; Logística e gerenciamento da cadeia de suprimentos. São Paulo: Atlas, 2015.

R CORE TEAM (2017); R: A language and environment for statistical computing. R Foundation for Statistical Computing, Vienna, Austria. URL https://www.R-project.org/.

ROACH, B.; Origin of the Economic Order Quantity formula; transcription or transformation? Managemente Decision, Vol. 43 Issue: 9, pp. 1262-1268, 2005.

TAHA, Hamdy A. Pesquisa operacional: uma visão geral. Tradução Arlete Simille Marques. Revisão Rodrigo Arnaldo Scarpel. 8. Ed. São Paulo: person Prentice Hall, 2008.

VENABLES, W. N.; SMITH, D. M.; the R Core Team. An introduction to R: Notes on R: A Programming Environment for Data Analysis and Graphics, 2016.